\documentclass[twocolumn]{aastex61}
\usepackage{graphicx,amsmath,amssymb,natbib,upgreek,bm}
\usepackage[greek,english]{babel}
\newcommand{\halpha}{\textrm{H\greektext a}}

\newcommand{\rmunit}{{\rm rad}\,{\rm m}^{-2}}
\shorttitle{A dense plasma globule in the solar neighborhood}
\shortauthors{Vedantham et al.}

\begin{document}

\title{A dense plasma globule in the solar neighborhood}

\correspondingauthor{H.K. Vedantham}
\email{harishv@caltech.edu}

\author{H.~K.~Vedantham}
\affiliation{Cahill Center for Astronomy and Astrophysics, MC 249-17, California Institute of Technology, Pasadena, CA 91125, USA}

\author{A.~G.~de Bruyn}
\altaffiliation{Deceased (2017 July 9$^{\rm th}$)}
\affiliation{Netherlands Institute for Radio Astronomy, Postbus 2, 7990 AA, Dwingeloo, The Netherlands}
\affiliation{Kapteyn Astronomical Institute, University of Groningen, Landleven 12, NL-9747 AD Groningen, the Netherlands}

\author{J.-P.~Macquart}
\affiliation{ICRAR/Curtin University, Curtin Institute of Radio Astronomy, Perth, WA 6845, Australia}

\begin{abstract}
The radio source J1819+3845 underwent a period of extreme interstellar scintillation between circa 1999 and 2007. The plasma structure responsible for this scintillation was determined to be just $1$-$3$\,pc from the solar system and to posses a density of $n_e\sim 10^2$\,cm$^{-3}$ that is three orders of magnitude higher than the ambient interstellar density \citep{ger2015}. 
Here we present radio-polarimetric images of the field towards J1819+3845 at wavelengths of 0.2, 0.92 and 2\,m. We detect an elliptical plasma globule of approximate size $1^\circ \times \gtrsim 2^\circ$ (major-axis position angle of $\approx -40^\circ$), via its  Faraday-rotation imprint ($\approx 15\,\rmunit$) on the diffuse Galactic synchrotron emission. The extreme scintillation of J1819+3845 was most likely caused at the turbulent boundary of the globule (J1819+3845 is currently occulted by the globule). The origin and precise nature of the globule remain unknown. Our observations are the first time plasma structures that likely cause extreme scintillation have been directly imaged.
\end{abstract}


\section{Introduction}  \label{sec:instr}
The light curves of a small fraction ($\sim 10^{-3}$\,epoch$^{-1}$) of extragalactic radio sources show the effects of passage through anomalously dense discrete interstellar plasma structures \citep{fiedler1987}.  Referred to as extreme scattering events (ESEs), these phenomena cannot be explained by the canonical Kolmogorov-like turbulence thought to pervade the interstellar medium \citep{armstrong1995}. Rather, they are inferred to be associated with small (0.1--10~AU) over-dense ($n_{\rm e}\sim $10--10$^3$~cm$^{-2}$) plasma structures \citep{fiedler1987}. How such over-dense plasma structures can exist in pressure equilibrium with the ambient ISM density of $n_e\sim 0.03$\,cm$^{-3}$ remains unresolved.

Several models have been advanced to eliminate the over-pressurization problem. One class of models invokes the ionized sheath of self-gravitating sub-stellar objects of neutral or molecular gas. These `failed stars' must, however, contain a significant fraction of the Galactic dark matter to account for the observed ESE rates \citep{walker1998}. 
Another class of models invokes chance alignments of plasma sheets extended along the line of sight. The sheets can furnish the required electron column at modest over-densities \citep{goldreich2006,pen2012}. Neither the elongated sheets, nor the isolated self-gravitating objects have been directly observed. 
%\citet{linsky2008} have proposed that short timescale (intra-hour such as J1819+3845) scintillations, in particular, are caused at the turbulent boundary between colliding local interstellar clouds within about 15\,pc from the Sun. 
More recently, \citet{walker2017} have proposed that extreme scintillation is caused by the photo-ionized sheaths of molecular clumps around hot stars, similar in nature to the cometary knots seen in planetary nebulae such as the Helix \citep[NGC7283;][]{vv1968}. 

The apparent impasse in determining the nature of these anomalous plasma structures is largely due to the lack of meaningful morphological information. Radio-wave scintillation only constrains transverse fluctuations in the {\em column} density of free electrons. It provides no meaningful information on the line of sight morphology of these anomalous plasma structures. The length-scale of transverse density perturbations probed by scintillation is given by the Fresnel length $r_{\rm F} \approx 10^{-4}(\lambda_{\rm cm}D_{\rm pc})^{1/2}\,{\rm AU}$, where $\lambda_{\rm cm}$ is the electromagnetic wavelength in cm, and $D_{\rm pc}$ is the distance to the scattering plane in parsecs. Scattering in the Galactic ISM therefore probes plasma density fluctuations on scales that are a tiny fraction of an AU. Evolution of scintillation properties over time yields some information on the spatial morphology of the implicated ISM structures on AU-scales but only over a quasi-linear transect. Hence, even the (two-dimensional) transverse morphology of these anomalous ISM structures has been inaccessible. 

Much of what we know about discrete plasma structures in the Galaxy comes from imaging of emission line nebulae such as HII regions which have typical emission measures of ${\rm EM}>10^4\,$pc\,cm$^{-6}$. The plasma structures implicated in radio wave scintillation typically have ${\rm EM}<1\,$pc\,cm$^{-6}$, and are therefore difficult to image via traditional means.

Here, we report the first direct imaging of an anomalous plasma structure implicated in  the scintillations of the radio source J1819+3845 whose extreme $\sim30$\%-modulated variations\footnote{Modulation index is the ratio of standard deviation of variations to the mean.}, exhibited on timescales $<0.5\,$hr, are attributed to turbulence estimated to be only $1$-$3$\,pc from Earth \citep{MdeB07,ger2015}.  The morphology of the plasma structure is revealed in images of its Faraday rotation imprint on the background diffuse polarized emission from the Galaxy, and enables direct comparison against proposed explanations of its physical properties. 

\section{Observations}
\subsection{Rotation measure synthesis}
The radio data presented here use the diffuse linearly polarized synchrotron emission from the Galaxy as a `back-light' to study intervening magneto-ionic plasma structures. Propagation through such plasma imparts a Faraday rotation of the electric field vector (EFV) by an angle given by $\phi = {\rm RM}\lambda^2$ where $\lambda$ is the wavelength of light and ${\rm RM}$ is the rotation measure.
The rotation in the phase angle of the complex polarization vector (CPV): ${\rm P}={\rm Q}+{\rm i}{\rm U}$, is $\chi = 2\phi$  \citep[see][for a review]{burn1966,brentjens2005}. 

We use the RM synthesis technique \citep[\texttt{rmsynthesis} hereafter;][]{brentjens2005} to convert spectral image cubes of the CPV to Faraday cubes. This step separates the polarized emission into components corresponding to different RM values. 
 
\subsection{WSRT 92-cm}
Data were acquired with the Westerbork Synthesis Radio Telescope (WSRT) in 2013 April on four consecutive nights (12\,h per night). We obtained visibilities on baselines that have separations of $b+72i$\,m ($i=0,1,2,...37$), where $b=36,\,54,\,72,\,$and\,$90$\,m for the four nights respectively. Visibilities had a contiguous frequency coverage between $311$ and $381$\,MHz, with a resolution of $156$\,kHz. Images were made at each channel with a $uv$-taper (25\% value) of 650\, 1300\, 2600\,m  baseline length at 350\,MHz. 

Most of the polarized Galactic emission appears between ${\rm RM}\approx 30$ and $65$\,rad\,m$^{-2}$, broadly consistent with the range of values measured for extragalactic sources in this field \citep{taylor2009,ger2015}. Figure \ref{fig:wsrt_lofar} (top-left panel) shows the peak polarized emission (across all RM values) in each pixel for the mid-resolution data (PSF $\approx 150\arcsec\times100\arcsec$). We find mottled emission across the field with a brightness of $\approx 1\,$K. The top-right panel of Fig.\,\ref{fig:wsrt_lofar} shows the Faraday image --- the RM at which the peak occurs. Here we see a discrete elliptical region of size $1^\circ\times\gtrsim 2^\circ$ (${\rm PA}\approx -40^\circ$) where the Faraday depth is offset by about $15\,\rmunit$ relative to the background. J1819+3845 is co-incident ($\lesssim$ few PSF-widths) with the Northern edge of the plasma globule. We argue in \S 3 that the scintillations were caused at the turbulent boundary of the globule. The scintillation data constrains the declination speed of the plasma to be about $+20$\,km\,s$^{-1}$ due North \citep{ger2015}; J1819+3845 is therefore currently being occulted by the globule.

\subsection{LOFAR 2-m polarimetry}
The 2\,m interferometric data were acquired on 2013 June 27 with the Low Frequency Array \citep[LOFAR;][]{vanhaarlem2013}. Visibilities were acquired over a 74\,MHz bandwidth centered on 152\,MHz. Interference was flagged at the correlator resolution of about 3\,kHz and 2\,s using the \texttt{aoflagger} program \citep{aoflagger}. The data were then averaged to a resolution of 195\,kHz and 10\,s. The 48 core-stations yield visibilities on baselines up to about 3\,km in length. We then used the BlackBoard SelfCal (BBS) LOFAR calibration package to calibrate towards, and peel the bright off-axis (about 20\,deg from J1819+3845) radio source Cygnus\,A  with a solution cadence of 195\,kHz and 1\,min. We used a two-component model to account for the dominant lobes of Cygnus\,A. 
We then constructed a model for the field using the 327\,MHz WENSS catalog \citep{wenss} and the 74\,MHz VLSS catalog \citep{vlss}. By using a nominal model for the LOFAR primary beam, we identified all sources present in both catalogs and had $>1\,$Jy of apparent flux density at 150\,MHz. We used this model to bandpass calibrate the visibilities at a solution cadence of 195\,kHz and 1\,min. We then subtracted these sources from the visibilities. Finally, we predicted the ionospheric Faraday rotation during the 8\,hr synthesis using \texttt{RMextract} software, and de-rotated the Stokes Q and U visibilities to remove this effect. 

\begin{figure*}
\centering
\begin{tabular}{ll}
\includegraphics[width=0.5\linewidth]{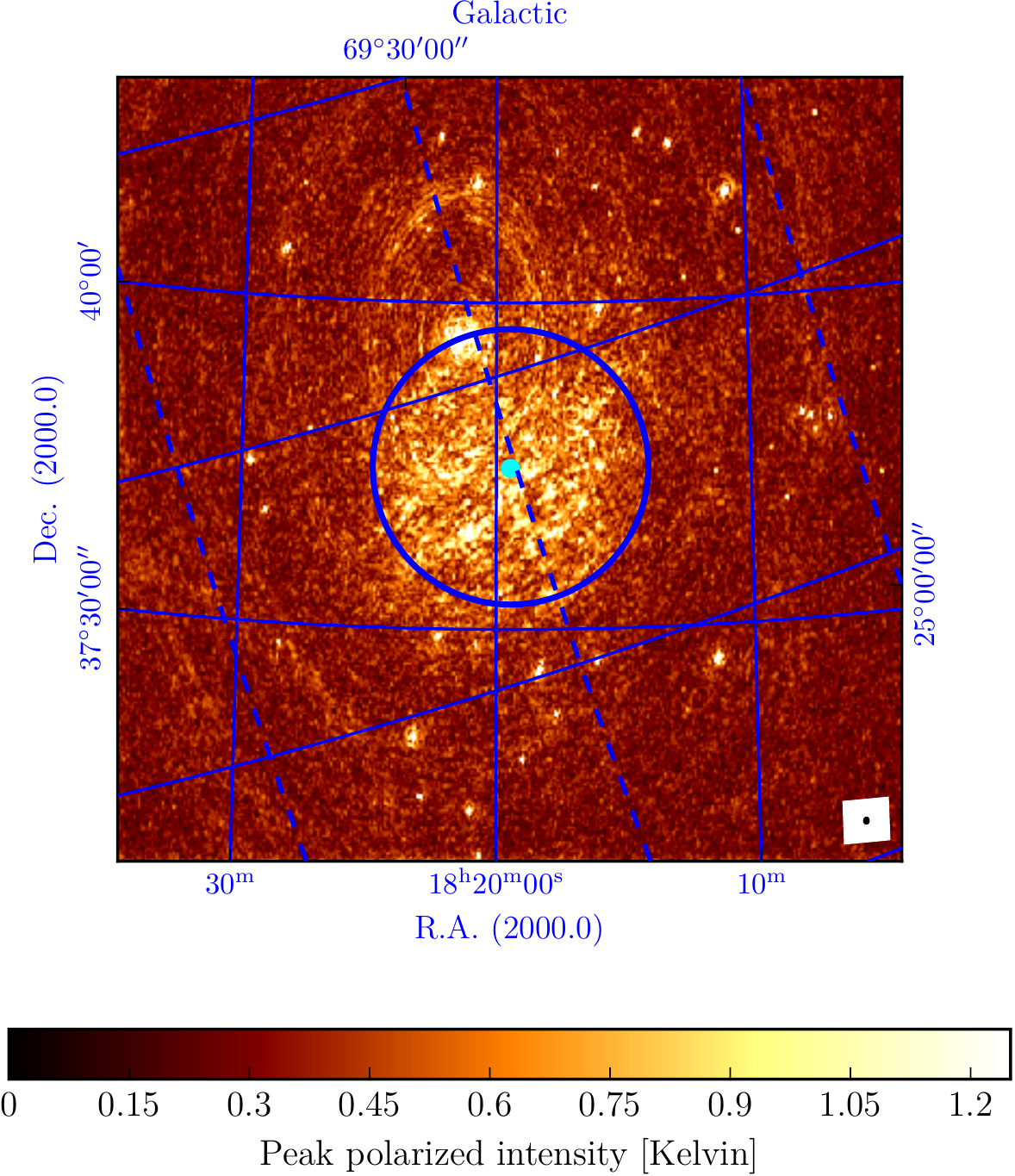} & \includegraphics[width=0.5\linewidth]{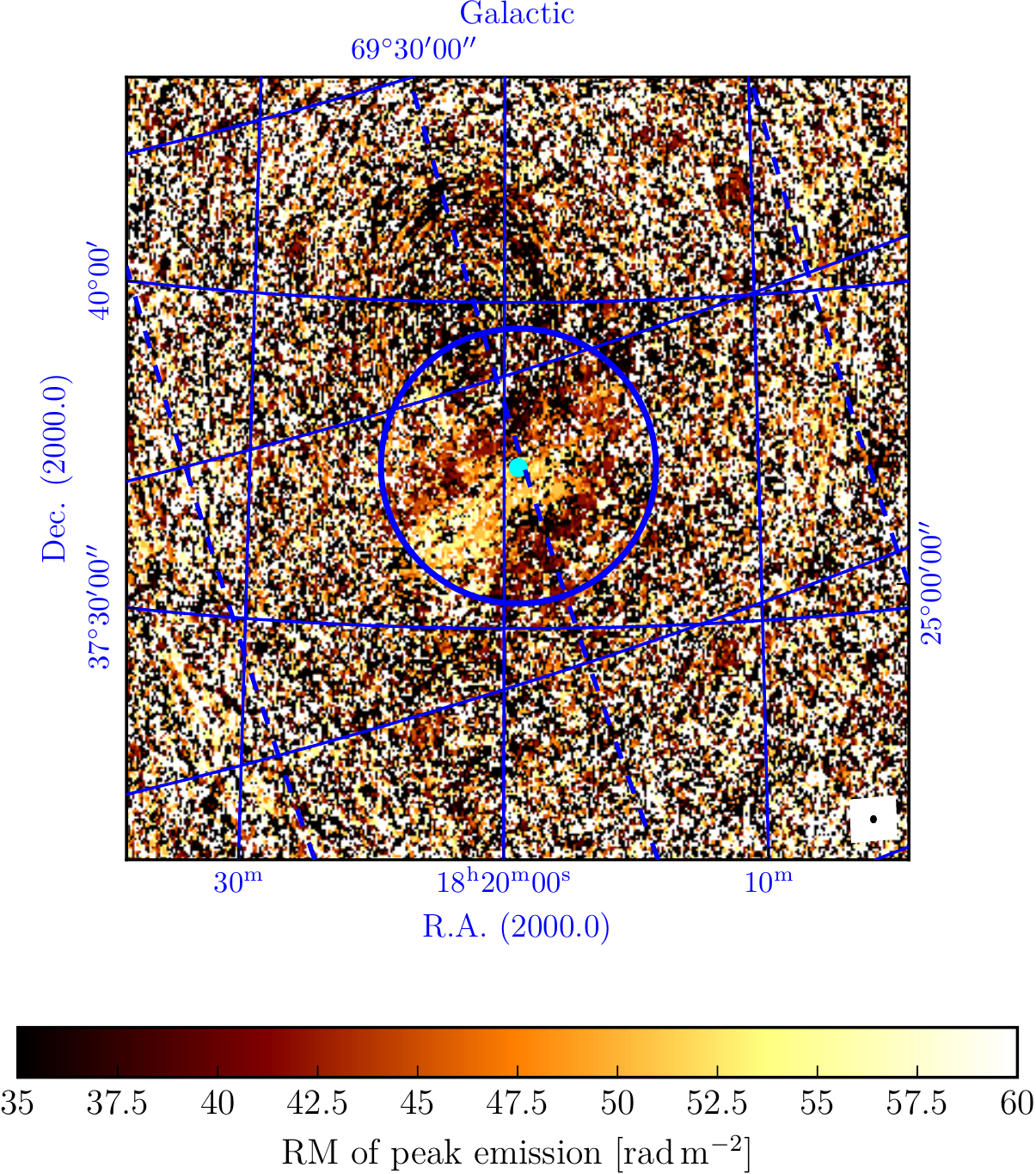}\\
\includegraphics[width=0.5\linewidth]{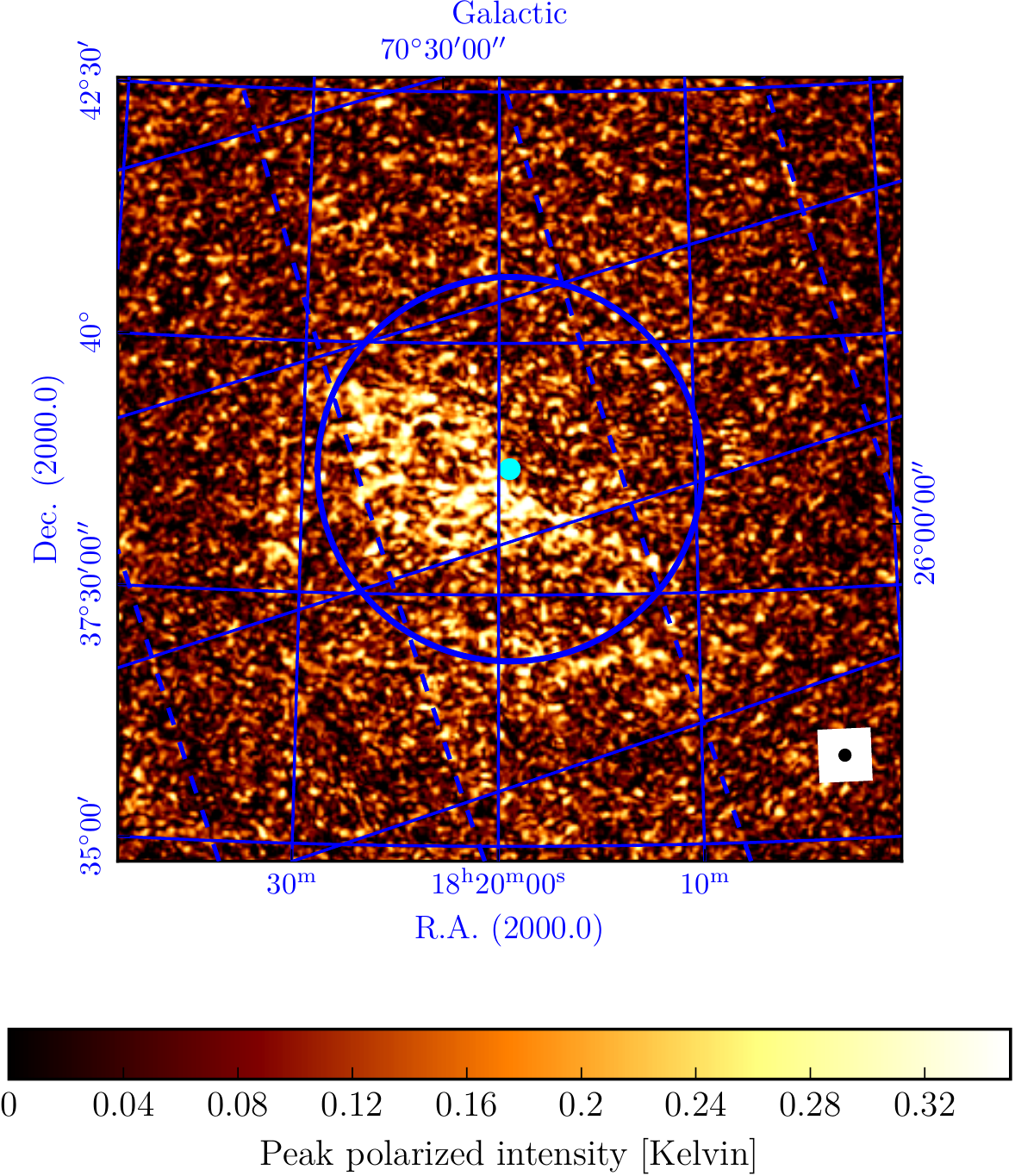} & \includegraphics[width=0.5\linewidth]{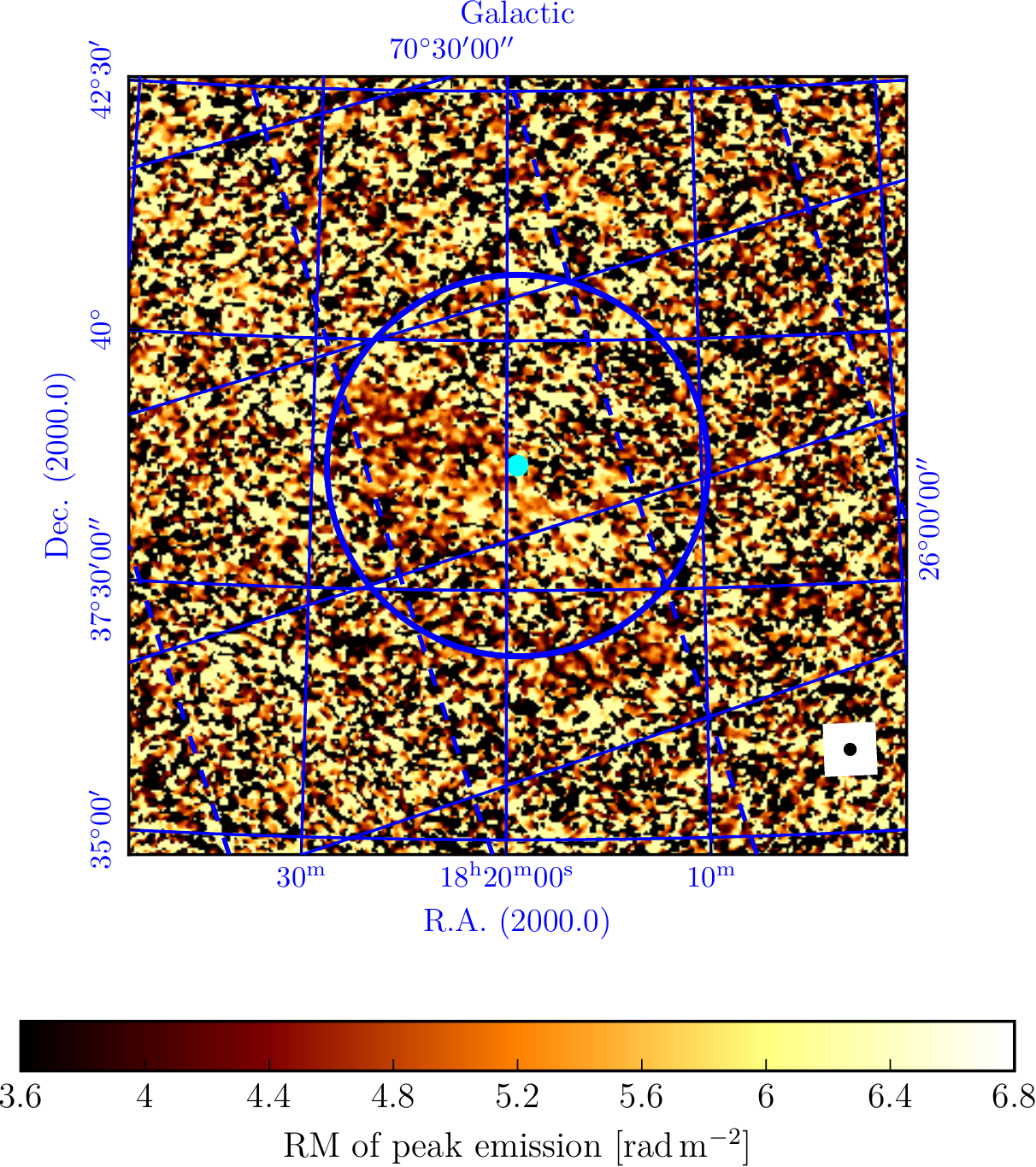}\\
\end{tabular}
\caption{Peak polarized intensity (left-panels) and rotation measure of the peak polarized intensity (right-panels). Upper row: WSRT 92\,cm data. Lower row: LOFAR 2\,m data. The bright rings centered on 18h21m20.8s, +39$^\circ$42$\arcmin$45$\arcsec$  in the 92-cm data are due to imperfect subtraction of the bright quasar 4C+39.56. Black ellipses in the insets (white background) show the beam size. The location of J1819+3845 is shown with a filled cyan circle. \label{fig:wsrt_lofar}}
\end{figure*}

Dirty stokes Q and U images at 195\,kHz wide channels were made using the {\em CASA} imager at $4\arcmin$, $6\arcmin$ and $8\arcmin$ resolution. Using \texttt{rmsynthesis}, we searched for polarized emission between RMs of $-100$ and $+100\,\rmunit$. We detected several polarized point-like sources known from the NVSS catalog at RMs ranging from about $50\,\rmunit$ to $100\,\rmunit$. Faint ($\sim 100\,$mK, in $6\arcmin$ images) diffuse emission was observed only at relatively low RMs of around $5\,\rmunit$ (see Figure  \ref{fig:wsrt_lofar}). We separately made Faraday cubes by splitting the available bandwidth into three parts. This diffuse emission is only seen in the upper band, suggesting significant beam and/or depth-depolarization of the emission at 2\,m wavelength. 

\subsection{WSRT 21-cm polarimetry}
The data processing of the 20-cm images is described in \citet{ger2015}. We find about 50\,mK of diffuse Galactic emission in $60\arcsec\times 80\arcsec$ images in the J1819+3845 field, centered at ${\rm RM}\approx 60\,\rmunit$ (image shown in Fig.\,\ref{fig:20cm_edge}).

\section{Discussion}
\subsection{Interpretation of the polarimetric maps}
The multi-wavelength data, with their differing sensitivity to diffuse structures and Faraday depths, enables us to assemble a comprehensive picture of the magneto-ionic medium towards J1819+3845.

Polarized Galactic emission of a few Kelvin brightness on angular scales of $\gtrsim 1\arcmin$ is ubiquitous in interferometric measurements made at $\lambda \approx 1$\,m \citep{gerAN}. The polarized structure is typically uncorrelated with total intensity structure \citep{bernardi1, bernardi2} because the bulk of the polarized structure is the result of spatial variations in the Faraday rotation in the intervening magneto-ionic medium, rather than structure intrinsic to the emission itself.  Based on the structure evident in the 92-cm images in Fig.\,\ref{fig:wsrt_lofar}, we argue that this applies to the field surrounding J1819+3845. 

The key to a unified interpretation of the three polarization datasets rests upon the fact that Faraday images made in each spectral window are sensitive only to RM fluctuations that lie in a `sweet spot' between two opposing factors. Only RM fluctuations  with sufficient amplitude on scales comparable to the interferometer fringe spacing are detectable. By contrast, RM fluctuations of sufficiently large amplitude on scales comparable to the synthesized beam lead to beam depolarization and hence non-detection.

We attempted to quantify the interplay between these effects in our data by performing a simple one-dimensional integration of the CPV over a range of phase angles to obtain the following approximate result: beam-depolarization reaches levels of 50\% and 90\% when the fluctuations in ${\rm RM}\lambda^2$ over the interferometer resolution reach levels of $0.6\pi$ and $0.9\pi$ respectively. Interferometric sensitivity peaks around ${\rm RM}\lambda^2 \approx 0.4\pi$, and reaches 50\% of its peak value at ${\rm RM}\lambda^2 \approx 0.1\pi$. We further note that the Faraday depth resolution, $\Delta {\rm RM}$ is limited by the wavelength coverage, $\Delta \lambda^2$, to values $\Delta \lambda^2 \, \Delta {\rm RM} \gtrsim 0.5$.

Our interpretation is as follows. The main feature of interest is the $15\,\rmunit$ globule detected in the 92-cm data. The polarized emission evident in the 2-m images originates in front of the globule. Bulk of the 20\,cm emission originates behind the globule. The `RM-edge' imparted by the globule at its boundary is possibly detected at 20\,cm. 

We first reconcile the 2-m data with the Faraday imprint of the $15\,\rmunit$ elliptical globule evident in the 92-cm data. The absence of the corresponding structure in the 2-m image over the entire range from $-100$ to $+100\,\rmunit$ suggests that the 2-m polarized emission at $\approx 5\,\rmunit$ and $\approx 100\,$mK brightness originates in front of the globule at $\lesssim 1\,$pc from the Sun. This is consistent with the general expectation that the observed polarized emission at longer wavelengths originates from closer distances (the so called `polarization horizon') due to a combination of beam and depth depolarization. \citet{sun2011} find a polarization horizon of 4\,kpc at $\lambda=6$\,cm, which corresponds to 3.6\,pc at $\lambda = 2\,$m. While this is only a rough upper-limit\footnote{Other authors find distances that several times larger \citep[e.g.][]{uyaniker2003}.}, it is consistent with the relatively local origin of the 2\,m emission in our model. In addition, any polarized emission originating behind the globule at 92\,cm would suffer catastrophic beam depolarization at 2\,m; a $>90$\% depolarization at 2\,m, for instance, requires fluctuations of about $>0.7\,\rmunit$ over the $6\arcmin$ beam size,  corresponding to $>0.3\rmunit$ fluctuations at the $2\arcmin$ resolution of the 92-cm images\footnote{We assume that the variance of fluctuations versus angular scale is a power law with an index of -1.65 \citep{bernardi1}.}. This is consistent with the magnitude of RM variations that would yield the mottled structure in the 92-cm images.

We next consider the 20\,cm data. Here the mottled $\Delta {\rm RM} \gtrsim 0.3\,\rmunit$ structure seen at 92\,cm only imparts a $\sim 0.02$\,rad rotation of the CPV at 20\,cm and, as anticipated, is absent from the 20\,cm map. The only imprint from the globule that the 20\,cm data is expected to be sensitive to is its edge, where we expect a $\delta{\rm RM}\approx 15\,\rmunit$ change to yield a perceptible shift in the CPV position angle of $\approx 68^\circ$. If the CPV shift happens within the beamwidth, we expect to see a `canal'-like feature along the globule edge \citep{haver04}. We find a canal-like feature to the North-East of J1819+3845 at a projected distance of about $190\arcsec$ at a position angle of about $-37^\circ$ from J1819+3845 (see Fig.\,\ref{fig:20cm_edge}; green line). This RM-edge however does not extend throughout the field of view of the 20\,cm images. In addition `canals' are quite ubiquitous in 21-cm maps. \citet{canals} find that the average separation between canals in 20-cm maps is $\sim 5\arcmin$ which is comparable to the uncertainty in the location of the globule edge. As such, we are unable to unambiguously detect the globule's edge on the 20-cm maps.

\subsection{Relationship between the globule and the scintillations}
While it is obviously infeasible to make an unambiguous association of the observed magneto-ionic RM structure (arcminute to degree scales), with the turbulence responsible for the scintillations in J1819+3845 (micro-arcsecond scales), three remarkable characteristics make the case for an association compelling:
\begin{itemize}
\item Polarimetric imaging with LOFAR at $\lambda=2\,$m typically reveals $5-30$\,K of polarized emission depending on the field \citep{iacobelli2013,jelic2014,jelic2015,eck2017}. We instead only find $0.1\,$K of emission towards J1819+3845, suggesting the presence of a particularly turbulent depolarizing structure, such as the globule, well within the polarization horizon at $\lambda=2\,$m. This places the globule in the same distance range as the scintillating screen towards J1819+3845.

\item Plasma globules are rare. 
The WENSS polarization survey detected only one similar structure over a field of 1000\,sq.deg.; this structure was ring-like \citep{schnitz07}, with a radius of $1.4^\circ$ and an RM change of $|\Delta {\rm RM}| = 8\,\rmunit$ \citep{haver03}. The probability of finding a globule in the $2^\circ\times2^\circ$ field imaged at 92-cm by chance is only 2\%.
\item The edge of the structure is located within a few beamwidths, $\lesssim 6 \arcmin$, of the position of J1819+3845.  This is remarkable given that its scintillations ceased circa 2006.8 (see \citet[][fig. 3 upper-panel]{ger2015}). The proximity of the cloud edge suggests a connection of this event with the passage of this cloud edge from the line of sight.  We estimate the probability that J1819$+$3845 should lie within $6 \arcmin$ of the edge of a globule by chance. This is determined by the filling factor of the edges of such globules, equivalent to the product of the the filling factor of globules with the the fraction of their area subtended by edges. Based on the globule detected here, the total filling factor of such edges is $0.2$\%.  The chance of coincidental association appears negligibly small.  
\end{itemize}
The line-of-sight and transverse positional coincidence of the globule with the scintillating screen leads us to conclude that the scintillations in J1819+3845 were most likely caused by the turbulent edge of the globule.

\subsection{Thickness and proper motion of the edge}

The duration of the scintillations in J1819+3845 can be used to determine the thickness of the turbulent edge and bulk proper motion of the globule.  Scintillations were discovered in 1999, but J1819+3845 may have exhibited scintillation as early as 1986 \citep{ger2015}.  We make the reasonable assumption that the scintillation velocity with respect to the solar barycenter of $\sim 35$\,km\,sec$^{-1}$ may be equated to the bulk motion of the globule. If the edge normal makes an angle $\alpha$ with respect to the scintillation velocity, then the cloud edge thickness, $\Delta\theta_{\rm edge}$, scintillation duration ($\Delta T$) and screen distance are related by
\begin{eqnarray}
\Delta\theta_{\rm edge} = 61 \arcsec \, \sin \alpha \left( \frac{v_{\rm screen}}{35\,{\rm km\,s}^{-1} } \right) \left( \frac{\Delta T}{8\,{\rm yr}} \right) \left( \frac{D_{\rm scr}}{1\,{\rm pc}} \right)^{-1}.
\end{eqnarray} 
If the scintillations existed in 1986, $\Delta\theta_{\rm edge}$ could be as large as $160 \arcsec \sin \alpha \, D_{\rm pc}^{-1}$, of the same order as the angular scale of the mottled structure in the 92-cm maps. The proper motion of the globule of $\sim 75\arcsec$\,decade$^{-1}$ is detectable with future observations and, in combination with the measured scintillation velocity, could independently corroborate the distance to the globule. 

\begin{figure*}
\centering
\begin{tabular}{ll}
\includegraphics[width=0.5\linewidth]{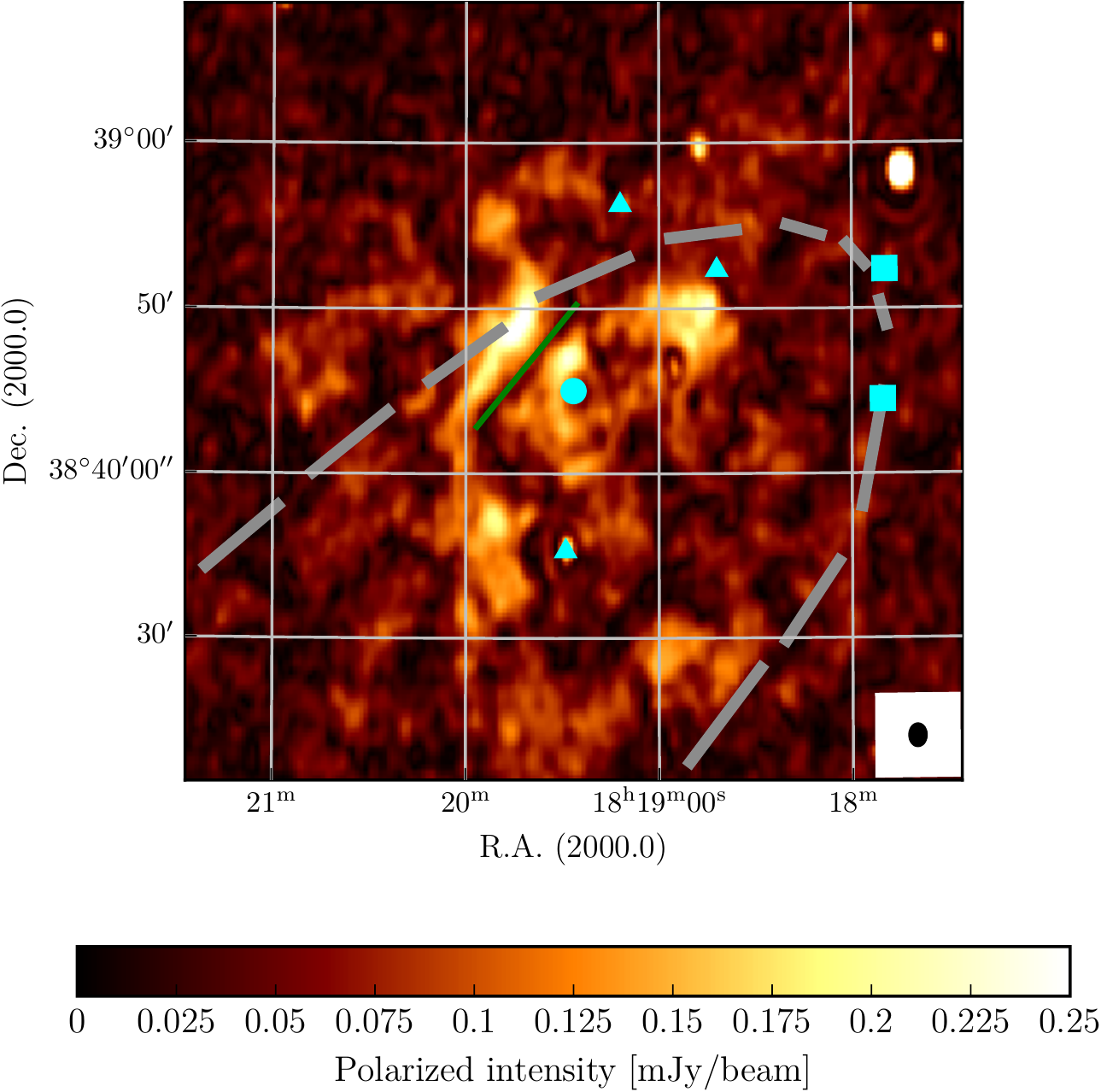} & \includegraphics[width=0.5\linewidth]{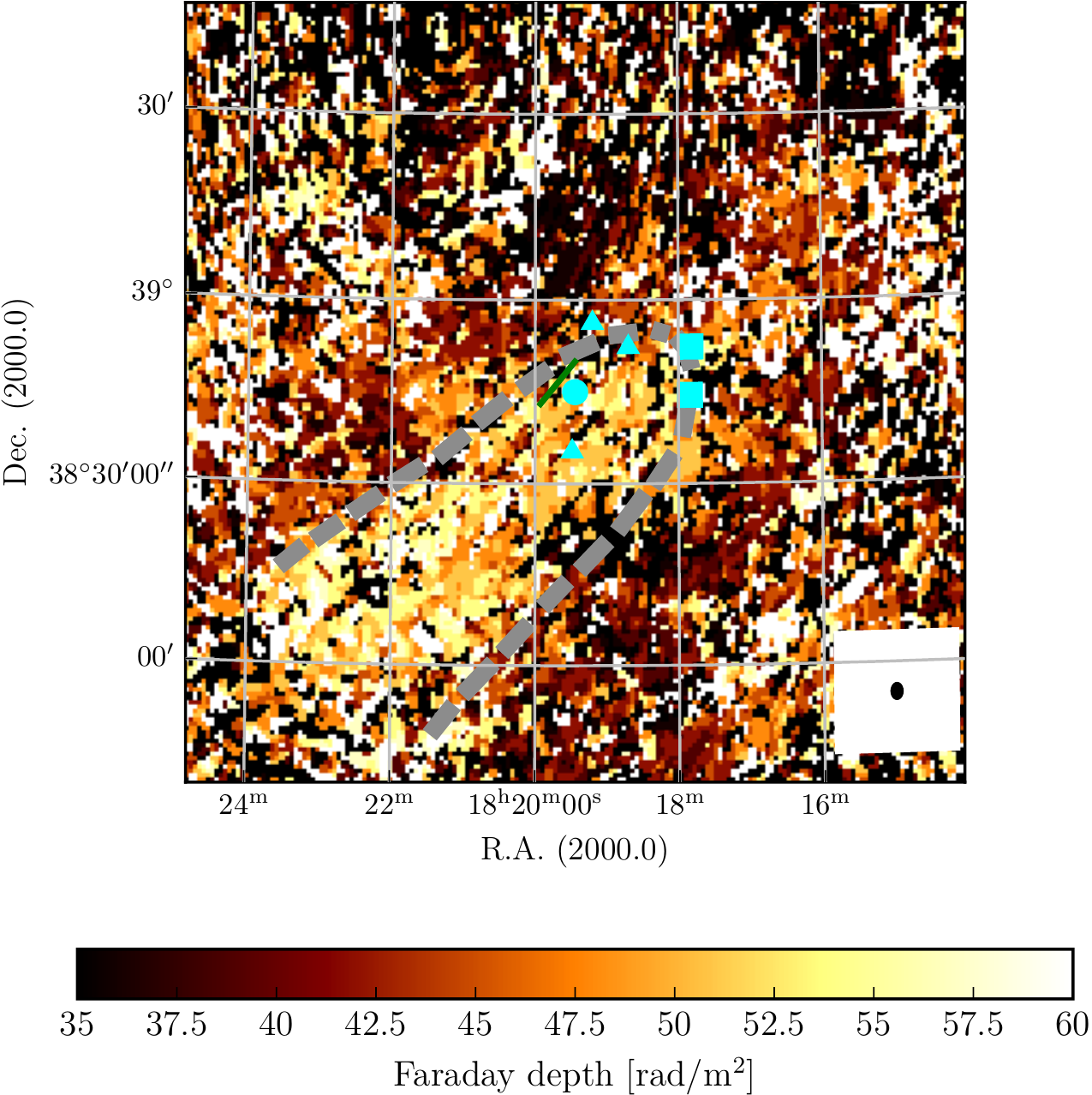} \\    
\end{tabular}
\caption{Left panel: 20-cm polarized intensity at ${\rm RM}=60\,\rmunit$. Right panel: Same as Fig. 1 top-right panel. Approximate contours of the globule seen in 92-cm data are overplotted in gray. The Faraday canal (see \S 3.1) is marked in green. Filled cyan circle shows the location of J1819+3845. Filled cyan triangles and squares respectively show the location of steep spectrum and flat spectrum sources whose light-curves are presented in \citet{ger2015}.\label{fig:20cm_edge} }
\end{figure*}

Fig. \ref{fig:20cm_edge} shows the locations of steep spectrum (cyan traingles) and flat spectrum (cyan rectangles) sources in the field for which \citet{ger2015} present light curves. Of these two source-classes, flat spectrum sources typically are compact enough to show scintillation \citep{masiv2}. Both flat-spectrum sources that lie close to the globule's boundary did exhibit  $\gtrsim 10$\% modulated flux variations as expected. Renewed monitoring of these and other flat-spectrum sources along the globule's edge can further constrain its thickness and proper motion.

\subsection{Physical properties of the globule}
Expressions governing the physical properties of the globule, incorporating the derived $(8.4 \pm 0.6) \times 10^{-3}\,$kpc\,m$^{-20/3}$ scattering measure, are given by \citet[][their \S 4.4]{ger2015}. Their main assumptions are (1) Kolmogorov-type turbulence, (2) equipartition between the thermal electrons and magnetic field (plasma $\beta=1$), and (3) an emission measure (EM) increment of $2\,$pc\,cm$^{-6}$.  We revisit these constraints in light of our work--- specifically, the degree-scale size of the globule and its ${\rm RM} \approx 15\,\rmunit$ offset. To account for the large angular scale of the globule, we smoothed the Virginia Tech \halpha~survey maps\footnote{Details and data may be found at \url{http://www1.phys.vt.edu/~halpha/}} \citep{vtss} to a resolution of $\approx 10\arcmin$ to arrive at a constraint on the \halpha brightness of the globule of about $1.5$\,Rayleigh (${\rm EM} \approx 4$\,Rayleigh). We further assume an electron temperature of $10^4\,$K, $D_{\rm src}=1\,$pc, and globule size of $2^\circ$ (i.e.\,a depth of $\Delta L \approx 7\times 10^4\,$au.).  The large RM of the globule can be reconciled if we relax the equipartition assumption; for $D_{\rm scr}=1$\,pc (or 3\,pc), a plasma $\beta$-parameter of $<1/7$ (or $<1/4$) is required, suggesting that the dense plasma is magnetically confined. The corresponding electron density and magnetic field strength are in the range $<10$ (or $6$)\,cm$^{-3}$ and $<50$ (or $30$)\,$\mu$G respectively, and the outer-scale of turbulence is $l_o>0.1\,$au. We caution that the region under consideration sits at the interface of the ISM and a dense globule, and thus in a highly dynamic environment. As such, the assumption of Kolmogorov turbulence as employed here may not apply.

\subsection{Outlook}
It is difficult to directly constrain the thin-sheets model of \citet{goldreich2006} \& \citet{pen2012} with our observations. The putative sheets probably possess a thickness of $\sim 10^{10}$\,cm \citep{goldreich2006} whereas our RM observations have a transverse resolution of $10^{15}\,D_{\rm src}$\,cm. A theoretical investigation into the feasibility of the development of such sheets at a turbulent boundary of a dense globule may prove fruitful. 

The self-gravitating `failed stars' model is difficult to reconcile with the observed globule. The proposed neutral objects have a size of $\sim 1$\,au. with an ionized sheath probably ten times as large \citep{walker1998}, whereas the globule we find is $\gtrsim 10^3$ times larger. However, we make the distinction between rapid intra-day variability observed in J1819+3845 and singular U-shaped time-symmetric events found by \citet{fiedler1987}; this model may still be applicable to the latter phenomenon and not to J1819+3845. 

The association of the RM globule with the proposed cometary knots around Vega \citep{walker2017} suffers from three discrepancies. (1) The likely distance to the globule is $1-3\,$pc, whereas Vega is $7.8$\,pc away. (2) The orientation of the RM globule (${\rm PA}\approx -40^\circ$) is not that expected for a cometary tail formed due to photo-ablation from Vega's radiation (Vega is at ${\rm PA}\approx 90^\circ$ from J1819+3845). (3) At a size of $10^3-10^4\,$au, the globule is far larger than known size of cometary knots in planetary nebulae. Regardless, the proposed analogy to cometary knots is only suggestive \citep{walker2017}, and we advocate for a search for molecular gas associated with the globule via imaging in ro-vibrational lines.

Finally, we remark that the globule may be the closest astronomical object to the solar system. The large proper motion of the globule is a falsifiable prediction of its association with the scintillations of J1819+3845.

\section*{acknowledgements}
The Westerbork Synthesis Radio Telescope is operated by the ASTRON (Netherlands Institute for Radio Astronomy) with support from the Netherlands Foundation for Scientific Research (NWO). This paper is based (in part) on data obtained with the International LOFAR Telescope (ILT). LOFAR (van Haarlem et al. 2013) is the Low Frequency Array designed and constructed by ASTRON. It has facilities in several countries, that are owned by various parties (each with their own funding sources), and that are collectively operated by the ILT foundation under a joint scientific policy. Virginia Tech Spectral-Line Survey (VTSS) is supported by the National Science Foundation. %The radio images were rendered using the \texttt{Kapteyn} package \citep{kapteynpackage} which uses \texttt{matplotlib} \citep{mpl} and Mark R. Calabretta's \texttt{wcslib} packages.


\begin{thebibliography}{}
\bibitem[Armstrong et al.(1995)]{armstrong1995} Armstrong, J.~W., Rickett, B.~J., \& Spangler, S.~R.\ 1995, \apj, 443, 209
\bibitem[Bernardi et al.(2009)]{bernardi1} Bernardi, G., de Bruyn, A.~G., Brentjens, M.~A., et al.\ 2009, \aap, 500, 965
\bibitem[Bernardi et al.(2010)]{bernardi2} Bernardi, G., de Bruyn, A.~G., Harker, G., et al.\ 2010, \aap, 522, A67
\bibitem[Brentjens \& de Bruyn(2005)]{brentjens2005} Brentjens, M.~A., \& de Bruyn, A.~G.\ 2005, \aap, 441, 1217
\bibitem[Burn(1966)]{burn1966} Burn, B.~J.\ 1966, \mnras, 133, 67
\bibitem[Cohen et al.(2007)]{vlss} Cohen, A.~S., Lane, W.~M., Cotton, W.~D., et al.\ 2007, \aj, 134, 1245 
\bibitem[de Bruyn et al.(2006)]{gerAN} de Bruyn, A.~G., Katgert, P., Haverkorn, M., \& Schnitzeler, D.~H.~F.~M.\ 2006, Astronomische Nachrichten, 327, 487
\bibitem[de Bruyn \& Macquart(2015)]{ger2015} de Bruyn, A.~G., \& Macquart, J.-P.\ 2015, A\&A, 574, A125
\bibitem[Dennison et al.(1998)]{vtss} Dennison, B., Simonetti, J.~H., \& Topasna, G.~A.\ 1998, \pasa, 15, 147
%\bibitem[Dennett-Thorpe \& de Bruyn(2002)]{dt2002} Dennett-Thorpe, J., \& de Bruyn, A.~G.\ 2002, Nature, 415, 57  
\bibitem[Fiedler et al.(1987)]{fiedler1987} Fiedler, R.~L., Dennison, B., Johnston, K.~J., \& Hewish, A.\ 1987, Nature, 326, 675
%\bibitem[Fiedler et al.(1994)]{fiedler1994} Fiedler, R., Dennison, B., Johnston, K.~J., Waltman, E.~B., \& Simon, R.~S.\ 1994, ApJ, 430, 581 
%\bibitem[Haslam et al.(1982)]{haslam} Haslam, C.~G.~T., Salter, C.~J., Stoffel, H., \& Wilson, W.~E.\ 1982, \aaps, 47, 1 
\bibitem[Haverkorn, (2003)]{haver03} Haverkorn, M., Katgert, P. \& de Bruyn, A.G. 2003, A\&A, 404, 233
\bibitem[Haverkorn et al.(2004)]{haver04} Haverkorn, M., Katgert, P., \& de Bruyn, A.~G.\ 2004, \aap, 427, 549
%\bibitem[Lang(1969)]{kenneth1969} Lang, K.~R.\ 1969, Science, 166, 1401 
\bibitem[Lovell et al.(2008)]{masiv2} Lovell, J.~E.~J., Rickett, B.~J., Macquart, J.-P., et al.\ 2008, \apj, 689, 108-126
\bibitem[Macquart \& de Bruyn(2007)]{MdeB07} Macquart, J.-P. \& de Bruyn, A.G. 2007, \mnras, 380, L20
\bibitem[Offringa et al.(2012)]{aoflagger} Offringa, A.~R., van de Gronde, J.~J., \& Roerdink, J.~B.~T.~M.\ 2012, A\&A, 539, A95
\bibitem[Pen \& King(2012)]{pen2012} Pen, U.-L., \& King, L.\ 2012, MNRAS, 421, L132
\bibitem[Rengelink et al.(1997)]{wenss} Rengelink, R.~B., Tang, Y., de Bruyn, A.~G., et al.\ 1997, \aaps, 124,
%\bibitem[Rickett(1969)]{rickett1969} Rickett, B.~J.\ 1969, Nature, 221, 158 
%\bibitem[Scheuer(1968)]{iss1968} Scheuer, P.~A.~G.\ 1968, Nature, 218, 920 
\bibitem[Shukurov \& Berkhuijsen(2003)]{canals} Shukurov, A., \& Berkhuijsen, E.~M.\ 2003, \mnras, 342, 496
\bibitem[Schnitzeler et al.(2007)]{schnitz07} Schnitzeler, D.H.F.M., Katgert, P., Haverkorn, M. \& de Bruyn, A.G. 2007, A\&A, 461, 963
%\bibitem[Stinebring et al.(2001)]{arcs} Stinebring, D.~R., McLaughlin, M.~A., Cordes, J.~M., et al.\ 2001, ApJL, 549, L97 
\bibitem[Taylor et al.(2009)]{taylor2009} Taylor, A.~R., Stil, J.~M., \& Sunstrum, C.\ 2009, \apj, 702, 1230
%\bibitem[Wagner \& Witzel(1995)]{idv} Wagner, S.~J., \& Witzel, A.\ 1995, ARA\&A, 33, 163
\bibitem[Uyaniker et al.(2003)]{uyaniker2003} Uyaniker, B., Landecker, T.~L., Gray, A.~D., \& Kothes, R.\ 2003, \apj, 585, 785 
\bibitem[Walker et al.(2017)]{walker2017} Walker, M., Tuntsov, A., Bignall, H., et al.\ 2017, arXiv:1705.00964
\bibitem[Walker \& Wardle(1998)]{walker1998} Walker, M., \& Wardle, M.\ 1998, ApJL, 498, L125
\bibitem[van Haarlem et al.(2013)]{vanhaarlem2013} van Haarlem, M.~P., Wise, M.~W., Gunst, A.~W., et al.\ 2013, A\&A, 556, A2
\bibitem[Vorontsov-Velyaminov(1968)]{vv1968} Vorontsov-Velyaminov, B.~A.\ 1968, Planetary Nebulae, 34, 256
\bibitem[Sun et al.(2011)]{sun2011} Sun, X.~H., Reich, W., Han, J.~L., et al.\ 2011, \aap, 527, A74
\bibitem[Goldreich \& Sridhar(2006)]{goldreich2006} Goldreich, P., \& Sridhar, S.\ 2006, \apjl, 640, L159
\bibitem[Jeli{\'c} et al.(2014)]{jelic2014} Jeli{\'c}, V., de Bruyn, A.~G., Mevius, M., et al.\ 2014, \aap, 568, A101
\bibitem[Jeli{\'c} et al.(2015)]{jelic2015} Jeli{\'c}, V., de Bruyn, A.~G., Pandey, V.~N., et al.\ 2015, \aap, 583, A137
\bibitem[Van Eck et al.(2017)]{eck2017} Van Eck, C.~L., Haverkorn, M., Alves, M.~I.~R., et al.\ 2017, \aap, 597, A98
\bibitem[Iacobelli et al.(2013)]{iacobelli2013} Iacobelli, M., Haverkorn, M., \& Katgert, P.\ 2013, \aap, 549, A56
\end{thebibliography}
\end{document}